# JITScanner: Just-in-Time Executable Page Check in the Linux Operating System

Pasquale Caporaso[1,3], Giuseppe Bianchi[2], Francesco Quaglia[1]

[1]DICII - University of Rome Tor Vergata

[2]DIE - University of Rome Tor Vergata

[3]CNIT Natl. Network Assessment and Monitoring Lab

April 2024

**Abstract**

Modern malware poses a severe threat to cybersecurity, continually evolving in sophistication. To combat this threat, researchers and security professionals continuously explore advanced techniques for malware detection and analysis. Dynamic analysis, a prevalent approach, offers advantages over static analysis by enabling observation of runtime behavior and detecting obfuscated or encrypted code used to evade detection. However, executing programs within a controlled environment can be resource-intensive, often necessitating compromises, such as limiting sandboxing to an initial period. In our article, we propose an alternative method for dynamic executable analysis: examining the presence of malicious signatures within executable virtual pages precisely when their current content, including any updates over time, is accessed for instruction fetching. Our solution, named JITScanner, is developed as a Linux-oriented package built upon a Loadable Kernel Module (LKM). It integrates a user-level component that communicates efficiently with the LKM using scalable multi-processor/core technology. JITScanner's effectiveness in detecting malware programs and its minimal intrusion in normal runtime scenarios have been extensively tested, with the experiment results detailed in this article. These experiments affirm the viability of our approach, showcasing JITScanner's capability to effectively identify malware while minimizing runtime overhead.

## 1 Introduction

The importance of services running on contemporary computing systems is growing significantly. There is an increasing demand for solutions that extend beyond simply analyzing code statically. These solutions should offer the capability to actively control actions that might be executed by potentially harmful programs, rather than relying solely on static analysis. This approach is highlighted in several literature studies [19, 13, 11].

Dynamic analysis, surpassing its static counterpart in various aspects, concentrates on observing the actual behavior of malware. It focuses on activities like network connections, file system access, and system calls. This method simplifies the identification of obfuscated or encrypted code, which is commonly employed by malware programs. Nevertheless, dynamic analysis utilizing sandboxing brings about a considerable overhead and has been proven to be detectable in specific configurations [5, 6, 29, 20].

In this article, we take a different approach and introduce JITScanner (Just-in-Time Scanner), a Linux-based solution. This solution is based on identifying, along wall-clock-time, appropriate moments for conducting actual checks on the content of a specific executable page. The primary aim is to circumvent the necessity of checking pages that, although associated with a program or an external library used by the program, are not actively present in RAM.

We use the term "materialized" to denote a virtual page that currently resides in a page frame within the RAM, accessible by the application in its address space. Pages that are not materialized will prompt a fault to be managed by the operating system if accessed. For instance, pages that are `mmap`-ed in a Posix operating system at a certain point in time might materialize later during the application's execution. By focusing on pages that are actively accessed for instruction fetching, we aim to bypass the expense of checking pages that the application does not utilize for storing machine instructions needed for execution.

From a technical perspective, JITScanner encounters a significant challenge with pages that are allocated for Write-Execute (WX) usage. Simply checking these pages at the time their content materializes and upon their initial use for executing machine instructions is inadequate for confirming that they will not harbor exploitable malicious signatures accessible to attackers. At the same time, it is crucial to note that WX pages hold substantial relevance in legitimate scenarios, particularly in supporting Just-in-Time (JIT) compiling within language-specific virtual machines [14].

To address this issue, our solution incorporates support for a security-focused state machine, managed at the kernel level. This "shadow" state machine operates in a way that the actual actions an application can perform on any WX (Write-Execute) page—such as writing or fetching instructions—are entirely logical. In particular, at any given moment, only one of the two possibilities (W or X) is allowed to happen without being traced by the operating system kernel.

This approach enables us to identify instances where an executable page is revisited by the CPU to fetch machine instructions after its content has been updated. Consequently, this allows us to recheck for the potential presence of malicious signatures following the modification of the page's content, i.e., the re-materialization of its content.

Importantly, this solution also addresses scenarios where an attacker exploits encrypted code that is later decrypted over time into an executable page. This particular scenario is critical and cannot be effectively handled by traditional static analysis techniques and tools and it presents a potential loophole for evading antivirus software that conducts dynamic analysis of active programs [7].

In our solution, we also provide the support for checking the page content upon an instruction fetch if a previously materialized non-executable page is requested to become executable via specific system calls. The same applies to executable pages that are also made writable and undergo updates over time. By incorporating this functionality, we ensure that whenever a non-executable page is dynamically altered to become executable or when executable pages are permitted to be writable and are modified, our system conducts checks on their content during instruction fetch. This approach helps maintain security by verifying the integrity of pages that undergo permission changes or content updates, preventing potential vulnerabilities that might arise due to these transitions

The architecture of JITScanner includes a subsystem to allow the communication of executable-page snapshots to user level demons, which can perform the page-content check asynchronously, off the critical path of the applications under control. This subsystem has been designed in order to enable the exploitation of parallelism in the machine, which is a common feature of nowadays multi-processor/multi-core chip-sets. Furthermore, the asynchronous page-content check offered by JITScanner can evolve along time by simply updating the logic (or the dataset of reference signatures) characterizing the user-level daemons activity. At the same time, the core kernel-level engine of JITScanner still enables the possibility to carry out the synchronous check of an executable page, which can be particularly useful for assessing the presence of specific malicious signatures. Addtionally, it embeds mechanisms for protecting JITScanner from Denial-of-Service (DoS) attacks.

We conducted comprehensive testing of JITScanner using diverse non-malicious workloads as well as a range of malware samples. Our tests demonstrated that the overhead introduced by JITScanner is practically negligible. Moreover, the results highlighted the effectiveness of JITScanner in detecting malicious software.

The remainder of this article is structured as follows. Related work is discussed in Section 2. JITScanner is presented in Section 3. Its experimental assessment is provided in Section 4. Usage scenarios and directions for further extensions are discussed in Section 5. Conclusions are drawn in Section 6.

## 2   Related Work

The literature on techniques for detecting malicious software is very ample. One trend consists in the exploitation of machine learning techniques in order to identify malware components in both traditional [22] and cloud oriented platforms [17]. Compared to all these techniques, our solution is essentially orthogonal, since we focus on a low-level service for deferring (or hopefully avoiding) the check of a given virtual page content along time, which can be implemented also according to machine learning techniques exploiting some knowledge base on malicious executable-page signatures.

Memory forensic analysis relies on memory images, such as VM (Virtual Machine) images, to deduce the potential existence of malicious software or applications [3, 28]. While certain techniques aim to reduce the number of memory checks required, this approach fundamentally differs from our proposal. Our approach is centered on bypassing checks on non-materialized pages, which are still present within the file systems or VM dumps analyzed by memory forensic techniques. Additionally, our solution addresses encrypted code that dynamically undergoes decryption at a granular level and is installed on an executable page within the address space. In contrast, memory forensics solutions typically conduct checks at a coarser level since events of interest, like API calls, lead to memory dumps that are not immediate. Consequently, the transient content of memory before the actual dump is not actively tracked within these solutions. The distinction lies in our ability to focus on non-materialized pages and handle dynamically decrypted code at a fine granularity, which sets our approach apart from traditional memory forensics methods.

Our approach fundamentally differs from any static checker, like ClamAV [1], because we specifically aim to avoid checking the entire image of a program. Instead, we provide the capability for dynamic checks at runtime, even accommodating WX (Write-Execute) pages loaded into an application's address space. While certain antivirus tools perform operations during runtime as well [15], they do not utilize the kernel-level technique that we introduce, nor do they possess the capability to manage the virtualization of access permissions to critical pages, such as WX pages, as we do through our shadow state machine. Recent studies have shown that common antivirus tools can be circumvented by software that intelligently utilizes these types of pages within an application's address space [7]. These attacks often involve materializing critical binary code into these pages at a later time, for instance, by decrypting code in such a page during runtime—a tactic that can be directly countered by leveraging our solution.

Indeed, the detection of malicious software often involves behavioral analysis of applications, achieved through methods such as intercepting library calls or system calls. Antivirus solutions employing dynamic analysis [15] and eBPF (Extended Berkeley Packet Filter) kernel-level solutions [4, 24] typically follow this paradigm. This form of behavioral analysis operates independently from (and can be combined with) our approach because it does not specifically target the identification of signatures within code hosted on executable pages, as our solution does. Furthermore, our approach factors in the dynamic decryption of code blocks, which is a distinct focus compared to the behavioral analysis employed by these solutions.

Various operating system security solutions aim to enhance the robustness in managing an application's address space. For instance, some approaches involve making user-level pages inaccessible for instruction fetching when operating in kernel mode, as seen in solutions like KPTI (Kernel Page-Table Isolation) [2]. However, these approaches remain distinct and separate from our solution. The primary focus of these methods is to mitigate risks associated with running kernel-level software. They aim to enhance security by reducing vulnerabilities that may arise within the kernel space. In contrast, our solution targets the reduction of risks that are primarily dependent on the user-level portion of the application. We concentrate on fortifying security measures concerning the user-level aspects, particularly in identifying and mitigating risks associated with executable pages and their content within the application's address space.

Our technique leverages a sophisticated handling of unused bits within the entries of an active program's page table. While similar approaches have been utilized by kernel-level solutions oriented towards performance improvements—such as transparently managing page contents for checkpointing [23] or in the development of distributed shared-memory layers [21]—our innovation lies in utilizing these available bits to construct a completely new shadow state machine. This state machine forms the backbone of our JITScanner technique, enabling us to dynamically manage and monitor the execution of code within an application's address space with enhanced security measures.

In comparison to the OmniUnpack kernel-level security solution [18], there are distinct differences in the methodology and capabilities of our JITScanner approach. OmniUnpack intercepts the initial attempt to fetch an instruction from a modified WX (Write-Execute) page and marks the page, but the actual verification occurs at a later time. This delay could potentially allow an attacker to rewrite the malicious page, concealing its actual content before the inspection takes place. Conversely, our JITScanner utilizes page snapshots, enabling us to track every transition of an executable page's content, particularly each time it is reused for instruction fetching. This granular monitoring provides a more comprehensive observation of page content transitions, enhancing our ability to detect potential malicious alterations. Moreover, while OmniUnpack does not handle purely executable pages, limiting its ability to trace such pages in conjunction with WX pages, JITScanner overcomes this limitation. We specifically designed JITScanner to trace both purely executable pages and WX pages, which is crucial in countering sophisticated attack techniques like ROP (Return Oriented Programming). ROP techniques exploit gadgets located anywhere within both executable and WX pages, and our comprehensive tracing capability helps prevent such exploits. Additionally, OmniUnpack's compatibility is limited to Windows XP and cannot be deployed on more recent Windows operating system releases. In contrast, our solution is designed to function with current Linux releases and aligns with the latest support for developing Linux kernel modules, providing compatibility and usability with modern Linux systems.

In comparison to the work presented in [27] for Windows XP, there are several fundamental differences between its kernel-level service and our proposal with JITScanner. The work in [27] introduces a kernel-level service that logs any page accessed for execution but not considered trusted, subsequently allowing analysis of the page content to determine the presence of critical machine instruction patterns. However, there are two significant distinctions from our proposal. Firstly, the [27] approach relies on a preliminary classification of non-malicious pages (e.g., library pages) to trigger the inspection of executable pages. In contrast, JITScanner supports the inspection of any executable page as soon as its content materializes in RAM and is accessed via an instruction fetch, without requiring prior classification. Secondly, like

Table 1: Comparison of JITScanner with literature solutions

|  | JITS | MAAR [22] | ClamAV [1] | Tracee [4] | Deep-Hook [17] | Falco [24] | Will. et al. [27] |
|---|---|---|---|---|---|---|---|
| Can be used on non-virtualized environment | ✓ | ✓ | ✓ | ✓ | ✗ | ✓ | ✓ |
| Allows for dynamic analysis | ✓ | ✓ | ✗ | ✓ | ✓ | ✓ | ✓ |
| Remains effective against packed malware | ✓ | ✓ | ✗ | ✓ | ✓ | ✓ | ✓ |
| Uses signatures check | ✓ | ✗ | ✓ | ✗ | ✓ | ✗ | ✓ |
| Reduces memory search ranges | ✓ | na | ✗ | na | ✗ | na | ✗ |
| Allows monitoring of multiple memory writes on executable pages | ✓ | na | ✗ | na | ✗ | na | ✗ |

OmniUnpack, the work in [27] introduces a memory management mechanism that excludes the possibility of fetching instructions from a writable page at page mapping time. However, once the page is accessed by an instruction fetch, it becomes regularly usable after a copy is logged on the hard drive. This mechanism lacks support for scenarios where executable code within a WX page undergoes multiple updates throughout the application's lifespan. Conversely, our proposal, JITScanner, addresses this scenario directly by leveraging the shadow state machine to manage WX permissions and page snapshots. This capability allows us to monitor and manage updates to code residing in specific WX pages efficiently and effectively.

Approaches that rely on running applications within an emulation environment, as outlined in [9], aim to utilize shadow memory to detect potential malicious alterations in an application's address space. However, emulation-based approaches might introduce some level of interference or non-transparency while running applications within the emulation environment. In contrast, our proposal manages executable pages through a fully transparent kernel-level service. By operating at the kernel level, our solution ensures a higher degree of transparency in managing executable pages within an application's address space. This transparency is vital as it allows for seamless and efficient monitoring and management of executable pages without introducing additional layers or interference that might affect the application's operation or introduce complexities into the system.

Table 1 summarizes the aspects discussed in this section, showing a comparison of our solution against the current state-of-the-art methods.

Finally, this article extends our work in [10], in particular via the extension of the use-cases that are employed in the experimental assessment. More precisely, we have considered different/wider sets of malware applications to assess the effectiveness of JITScanner. Furthermore, we studied the case of applications based on Just-in-Time compiling technology—and relying on multiple differentiated programming languages—which can lead to stress the behaviour of JITScanner in terms of management of pages associated with WX permissions. This leads to enlarge the assessment of the performance impact of JITScanner. Furthermore, this article also deepens some methodological and technical aspects characterizing our solution.

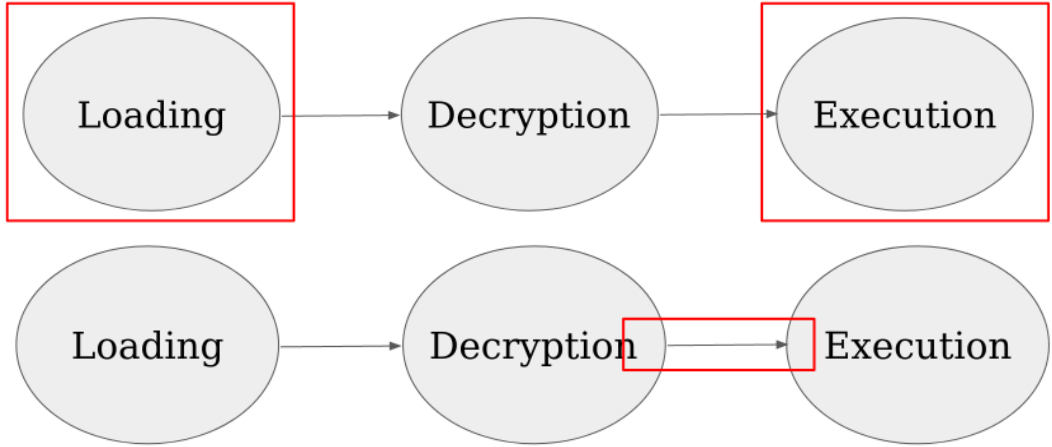

Figure 1: Common existing solutions (top picture) vs JITScanner (bottom picture)

## 3 The JITScanner System

### 3.1 Baseline Concepts and Methodology

The design principle of JITScanner is simple yet effective: we perform a binary code security check on an executable page *only when that page content is actually materialized in memory and is accessed for an instruction fetch*. This approach detects malware that decrypts malicious stages during the execution, and thwarts malicious code which delays execution to evade sandboxing.

At a high level of description, Figure 1 allows outlining the central difference between JITScanner and literature solutions for malware detection. In particular, the common way of proceeding in the literature consists of working at either the application loading time (e.g. with the static analysis of software and/or with the instrumentation via hooks—like in antivirus products) or the application runtime (e.g. via the interception of calls to services, like system-library or system-call usage). However, between these two phases malware software can also perform decryption steps of its own code during its execution, which is not directly interceptable by these load/execution steps based solutions. In JITScanner, we have the possibility to check any piece of code in an executable page just when execution on that page (re)-starts, possibly after a dynamic modification of the page content. Hence, from the actual usage of pieces of code fetched by the CPU for their execution, we enable taking snapshots of the code image in an executable page for their analysis. Overall, our approach ultimately combines dynamic with static analysis, as code (a snapshot of it) is analyzed at the time of its actual execution in a just-in-time fashion.

The malware detection conducted by JITScanner is based upon basic principles, specifically the reliance on page-fault handlers for intercepting and managing the CPU fetch of instructions from executable pages. Although JITScanner is the first solution to leverage this mechanism for the purpose of conducting memory analysis on Linux systems—while also including WX pages—the ideas it relies on are essentially reusable in any other operating system family. Overall, the methodology JITScanner exploits lies naturally on contemporary address space management solutions in common software systems. We believe this widens the applicability of our ideas.

In the following sections, we delve into the details of the architecture and technical approaches that underpin the design of JITScanner.

## 3.2 Architectural Hints

The objectives of our system are two-fold: we want to perform a comprehensive analysis of every executable page that can be exploited by an execution flow of a program, while minimizing at the same time the intrusiveness. To achieve these objectives, our system employs a multi-faceted approach, whose scheme is summarized in Figure 2.

The architecture of our solution exploits both a kernel-level layer and a user-level layer. The kernel-level layer, which is fully implemented as a Linux Kernel Module (LKM), performs two critical actions:

1) it intercepts the first access to an executable page (or an updated executable page—this is the case of WX pages), which is carried out via an instruction fetch by the CPU;

2) while handling the interception of the instruction fetch, and after the kernel materialized the target page in RAM, we perform critical checks on the page content which might lead to the forced termination of the application.

The verification process described in point 2) occurs within the same thread that initiated the access (for instruction fetch) to the executable page. Consequently, it operates synchronously concerning the application's execution.

To ensure swift handling of this scenario, allowing the thread to quickly resume executing user-level code if the application termination is not prompted, our architecture integrates modules responsible for conducting these critical checks directly at the kernel level. However, in typical situations, it is essential to restrict the number of checks performed synchronously. For instance, a common scenario might involve merely verifying if the page content includes the embedding of shellcode. Limiting the synchronous checks helps maintain efficient and responsive execution of the application by balancing the critical verification tasks without impeding the performance of the user-level code.

To still enable a more ample check of the page content, we exploit the user-level part of the architecture. It involves a program that receives input from the kernel in the form of a tuple: $<page\_content, offset, process\_id, thread\_id>$. The $offset$ denotes the position (linear address) of the first fetched instruction within the page. This program conducts a more comprehensive check of the page content asynchronously concerning the thread that initiated the page access. This approach allows for a broader examination to detect potential malicious signatures within the executable page, carried out off the critical path of the original application's execution. This checking program can be configured to take action, such as blocking the application (e.g., via the `kill()` system call), if it identifies any security-critical aspect during the asynchronous check.

Moreover, the *page_content* included in the tuple represents a snapshot of the executable page captured at the time of the intercepted instruction fetch. This snapshot capability enables the analysis of the complete history of the content of any WX (Write-Execute) page over time, a feature not facilitated by solutions like [18]. This historical analysis provides a more comprehensive view of changes occurring within WX pages, enhancing the detection and understanding of potential security threats.

The combination of an initial synchronous check at the kernel level followed by a subsequent asynchronous check at the user level provides configurability based on the criticality of the hosted services within the system. For services of lower criticality, the synchronous check can be omitted entirely. This prioritizes application performance, allowing the application to continue functioning without delay, even if the asynchronous check later identifies a malicious signature in an executable page that the application has already accessed. This modular approach offers flexibility and adaptability. It allows for straightforward customization and future expansion while enhancing reliability and ease of maintenance. By decoupling the synchronous and asynchronous checks and tailoring their execution based on service criticality, the system can strike a balance between performance optimization and security enforcement, ensuring efficient operation while maintaining a level of security appropriate for the service's importance.

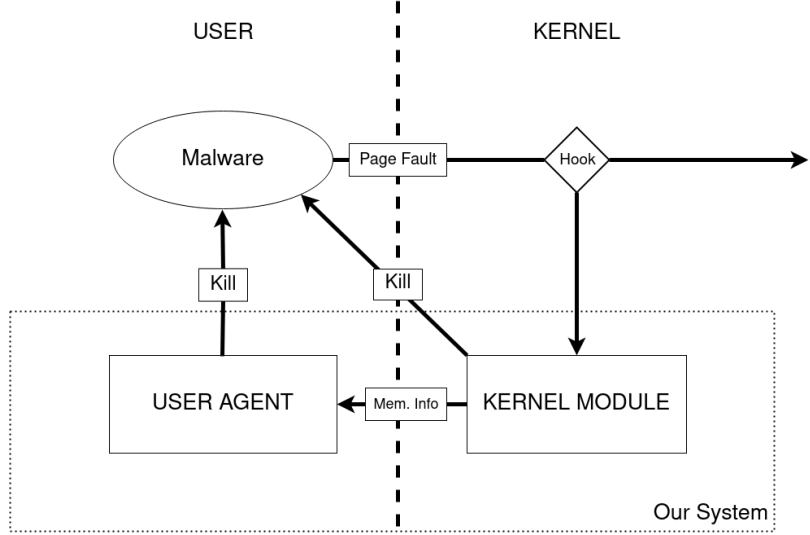

Figure 2: System Architecture

We also note that performing the kill of the malicious application in JITScanner is a baseline choice, but alternative solutions for the handling of the malware program can be simply adopted. As an example, there are works that have proposed the dynamic setup of facilities at the operating system level in order to manage malware suspicions in specific programs (see, e.g., [11]). The combination of JITScanner with these solutions can be immediate since, rather than killing the identified malware, JITScanner could simply alert an external system (passing to it baseline information, like for example identifiers of processes and users) in order to let that system start the management of the malware according to its own rules.

In the JITScanner architecture, the user-level component can effectively utilize existing code analysis methods accessible in the literature. In our current implementation, we have developed a user-level layer centered around Yara rules, a well-established system renowned for malware classification [26]. Regarding the kernel-level synchronous check, we have incorporated a baseline check to identify the presence of the meterpreter shellcode within the memory page.

In any case, the check facilities within our solution can be perceived as plug-ins that are highly extensible and can be effortlessly expanded at any given moment. Fundamentally, our solution comprises an engine dedicated to the management of executable pages within an address space. This engine serves as a stable core foundation upon which various checking mechanisms can be built, refined, and enhanced over time.

This modular architecture aligns with our primary research objective, which revolves around providing an effective kernel-level system to combat dynamic loading and obfuscation techniques. These techniques often pose significant challenges to traditional static analysis methodologies. By focusing on the core engine for managing executable pages and offering a flexible framework for diverse checking mechanisms, our solution aims to address these challenges and ensure an evolving and robust defense against dynamic loading and obfuscation techniques commonly utilized by malicious actors.

Our implementation has been tailored for kernel version 5 of Linux, and for x86 processors with PAE (Physical Address Extension) enabled. However, generalizing our LKM to support other architectures should be easy since we have utilized generic kernel functions wherever possible.

## 3.3 Executable-Page Access Interception

Within JITScanner's framework, the primary focus lies in intercepting a genuine access to an executable virtual page, particularly when the CPU actively fetches an instruction from that specific virtual page. Consequently, our solution does not initiate checks on any executable virtual page until the moment when the application actually requires the binary code contained within it.

In line with the typical on-demand paging process, this need arises after an executable page is allocated by the kernel and placed into a RAM frame. Subsequently, the page-table of the application is updated, granting the execution capability and making the page accessible. It is precisely at this point—when the CPU fetches instructions from the executable page—that JITScanner intervenes to conduct the necessary checks and verifications, ensuring the integrity and security of the fetched binary code.

To precisely identify the moment when a memory page is initially accessed for an instruction fetch, our solution employs a series of strategically placed kernel probes within the page fault handling process. Specifically, we utilize the `kprobe` subsystem provided by the Linux operating system to set up a hook on the return of a core page-fault handling procedure. In greater detail, we have installed a `kretprobe` on the `handle_mm_fault()` procedure within the Linux kernel. Upon completion of this procedure's execution, the target page that triggered the fault has already been materialized in RAM. At this stage, we can conduct the synchronous check on its content since it is now available in memory. Additionally, we take a snapshot of the page, which is managed for subsequent examination by the asynchronous check. This approach allows us to precisely time the execution of checks and efficiently manage the content of accessed pages for both synchronous and asynchronous analysis.

An essential consideration in our hook execution is to perform the page check only if the accessed page for instruction fetch is a legitimate page within the address space. Any attempt to access a non-legitimate page doesn't result in the actual materialization of the page in RAM through the operating system's `handle_mm_fault()` function. Instead, it leads to the delivery of the `SIGSEGV` signal to the application, indicating a segmentation fault or an invalid memory access attempt.

This distinction is particularly pertinent in our management of WX (Write-Execute) pages. For these pages, an "original" page fault triggered by the operating system—associated with the first access to the virtual executable page, which we intercept using the hook—is insufficient for checking the absence of malicious signatures. As a result, careful consideration is necessary to ensure that our page checks are executed only when the accessed pages are legitimate within the application's address space, thereby optimizing the effectiveness of our security measures while avoiding unnecessary checks on non-legitimate pages

For managing WX (Write-Execute) pages, we've developed an innovative kernel-level mechanism that dynamically adjusts the actual page permissions while the program is executing. Despite these dynamic modifications, the page remains "logically" accessible at the application level in both Write (W) and Execute (X) modes.

This kernel-level facility operates entirely transparently to the actual program, ensuring that the application is unaware of these permission modifications. However, it grants us the capability to monitor and track page access—whether for writing data or fetching instructions—throughout the program's execution.

Essentially, this facility establishes a shadow state machine that aligns with the legitimate permissions granted to the application for the page. By dynamically managing and tracking the permissions without the application's knowledge, we can maintain security while ensuring necessary access to the pages,

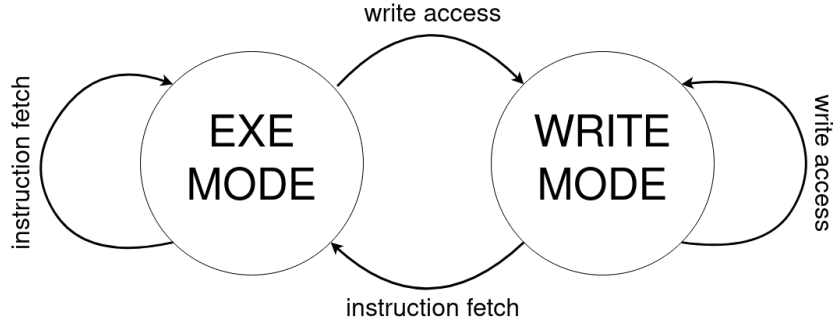

Figure 3: Shadow state machine for WX permissions

offering a nuanced and transparent approach to managing WX pages.

The shadow state machine within our system implements a time-separated alternating mechanism between write and execute permissions for pages. When a page is in "write mode" in the shadow state machine, it allows page-update operations without triggering user execution-mode interruptions such as page faults or interception via our kernel-level support. However, any attempt to execute instructions from that page will result in a fault, redirecting control to our kernel-level hook for interception and handling. Conversely, when a page is in "exec mode" within the shadow state machine, it permits instruction-fetch operations. In this mode, attempts to update the page will trigger a fault, intercepted by our kernel-level support for handling. This alternating logic between write and execute permissions ensures a controlled and secure mechanism for managing page access and operations within the application's address space. A representation of this logic is shown in Figure 3.

With our facility that manages page permissions dynamically, we have the capability to trigger the page-verification process whenever the access mode transitions between write and execute permissions. However, based on the goals of our solution, we believe that conducting a signature check on the executable page is essential only during instruction fetches. This is the critical moment where potential exploitation of malicious code could occur, making it the prime instance for initiating the signature check.

By specifically targeting the instruction-fetch operations for page verification, we can efficiently focus our security measures on the most vulnerable points in time, ensuring that checks for malicious signatures are conducted precisely when the executable page is being accessed for instruction execution. This targeted approach enhances the effectiveness of our security measures while optimizing performance by avoiding unnecessary checks during other access modes.

To support shadowing, we set the XD bit on the lowest level of the page table (the PTE), disabling instruction fetches from the page as soon as the WX page is materialized. By doing this, when the user program tries to execute the instructions on the page, a page fault occurs due to the lack of required permissions. We intercept this page fault and do the following:

- clear the XD bit, to allow the user to execute the page from now on;
- clear the W bit, to disable writes on the page;
- suppress the `SIGSEGV` signal generated by the invalid access to prevent the kernel from terminating the user program or making it run a `SIGSEGV` handler.

To handle the interception of kernel-level functions responsible for delivering the `SIGSEGV` signal during CPU fetches, our approach involves suppressing this signal since the page fault is generated by our shadowing mechanism. In Linux, faults occurring during instruction fetches are managed within the architecture-specific code of the kernel before calling the `handle_mm_fault()` function. However, these functions cannot be directly hooked using the `kprobe` mechanism. As an alternative, we have chosen to place a kernel probe on the `force_sig_fault()` function, which is triggered whenever an invalid memory access occurs. By hooking into this function, we can effectively suppress the delivery of the signal. Moreover, the synchronous checking of page content and the passage of page snapshots to the asynchronous checker are handled through the hook installed on the return of the `handle_mm_fault()` function. This comprehensive approach enables us to intercept and manage the necessary signals and functions, ensuring the seamless operation of our shadowing mechanism and the subsequent verification processes.

To effectively handle the hook for the `force_sig_fault()` kernel function, it is crucial to differentiate between standard invalid-access faults and those triggered by our shadow state machine. For this purpose, we utilize two unused bits within the lowest level page table entry (PTE) of the `x86` processor. These bits store the original write and execute permissions for a specific page before any modifications by our module.

During the interception of a fault, we check these bits within the PTE. If the fault originates from a genuine invalid access generated by the program and not by our shadow state machine, the bits help us identify this distinction. In such cases, we allow the kernel to handle the fault without any intervention from our module. This mechanism ensures that our system effectively identifies and manages faults triggered by the shadow state machine while allowing standard invalid-access faults to be handled conventionally by the kernel.

We present in Listings 1 and 2 the main parts related to the hook-based page management process. In Listing 1, we illustrate the management of our state machine in the `handle_mm_fault()` hook, which is invoked on page materialization and a write access violation. The steps are as follows: first, we do some general checks to ensure that the fault occurred in a valid `vm_area` with the appropriate flags. We then examine our `ORIG_WRITE_BIT` bit in the page table, to verify if the page has been modified by our state machine and if writes were initially permitted, if so, we disable execution and allow the write. Otherwise, we let the kernel handle the fault conventionally. In Listings 2, we report the hook for `force_sig_fault()`, which is called on an execute-address violation. Here we check if our `ORIG_EXE_BIT` bit is set in the page table. If it is not set, we let the kernel handle the fault. If it is set, this fault is generated under our state machine and needs to be resolved. To this end, we first run a synchronous check of the page content at kernel level. If a threat is detected, we allow the kernel to complete the fault, resulting in the delivery of the `SIGKILL` signal to the running process. If no threat is found during the kernel-level check, we transfer the page snapshot to the user agent for asynchronous checking and then fix the page table to allow the execution to continue.

The necessity of a TLB (Translation Lookaside Buffer) flush within the code listings is crucial due to the way TLBs maintain information based on the previous access permissions of a page. When the permissions of a specific page are modified, it becomes imperative to eliminate this page's address from the TLB across all CPUs involved in executing threads operating within the same address space as the intercepted thread accessing memory.

In our implementation, we've leveraged the `__flush_tlb_one_user` API provided by the Linux kernel for this purpose. This API allows us to perform a TLB flush for a specific page address. Additionally, we combine this API with the Inter-Processor-Interrupt (IPI) approach to execute the function across all CPUs involved. This approach ensures that the TLBs across all relevant CPUs are updated and purged of the outdated page information, aligning them with the modified access permissions for the page.

Listing 1: The hook on `handle_mm_fault()`
```
if (general_checks() == OK &&
    pte[fault_page] & ORIG_WRITE_BIT) {

    set_bit(fault_page, XD_BIT);
    set_bit(fault_page, W_BIT);
    set_bit(fault_page, ORIG_EXE_BIT);
    flush_tlb(fault_page);
    goto allow_write;
} else {
    goto standard_fault_handler;
}
```

Listing 2: The hook on `force_sig_fault()`
```
if (pte[rip_page] & ORIG_EXE_BIT) {
    if (sync_check(rip_page) != OK) goto send_sig_kill;

    transfer_page_to_user(rip_page);

    clear_bit(rip_page, XD_BIT);
    clear_bit(rip_page, W_BIT);
    set_bit(fault_page, ORIG_WRITE_BIT);
    flush_tlb(fault_page);
    goto allow_normal_execution;
} else {
    goto standard_fault_handler;
}
```

Finally, installing hooks on system calls like `mprotect()` is a significant step within our system. This particular system call is capable of altering the access permissions to a page, such as adding executable (X) or writable (W) permissions to a page that previously had different permissions. By placing hooks on system calls like `mprotect()`, we ensure the correlation between any change in access permissions and the need to recheck the page content upon the next instruction fetch from that page. This approach guarantees that whenever there is a modification in permissions using `mprotect()` or similar calls, our system is aware of it. Consequently, it can plan to perform necessary content rechecks when the page is accessed for instruction fetching, aligning with the modified permissions and ensuring security integrity.

## 3.4 Protection Against DoS

It is integral for our system that the page fault hook and the driver, responsible for transferring page content and metadata to the user-space agent, communicate efficiently. In our architecture, we employ a hash table for this purpose. This hash table organizes intercepted pages, creating copies (page snapshots) and categorizing them based on their associated process ID. These pages are stored within different buckets in the hash table, allowing parallel access for both insertion (storing a copy of an executable page to be checked) and deletion (transferring the page copy to a user-space buffer).

However, a critical aspect of this solution is the potential for a flow of page-fault interceptions, potentially caused by a malicious program intending to disrupt our page-management system. This influx of page faults can result in uncontrolled effects due to the intensive memory usage required to store copies of executable pages in the kernel-level hash table before transferring them to the user-space agent.

This scenario of numerous intercepted page faults can lead to an excessive volume of memory usage within the kernel, potentially impacting system stability and performance. It is crucial to develop mechanisms or safeguards within the system to handle such scenarios efficiently, ensuring that the system remains resilient against potential attacks while maintaining its stability and operational integrity.

To fortify the system against Denial of Service (DoS) attacks stemming from memory depletion due to excessive kernel-level allocations, we have integrated an additional hash table. This secondary hash table maintains support for parallel access across different buckets. Each entry within this hash table is responsible for tracking the user ID associated with pending executable-page copies awaiting delivery to the user-level agent, along with a counter representing these pages.

As a preventive measure, we have implemented a threshold mechanism within the system. When the counter surpasses a specified threshold value for a particular user's pending page copies, it signifies an abnormal influx of page faults. In response, actions are taken to prevent potential DoS scenarios. The system can be configured to either terminate or block the process generating additional page faults on behalf of the same user.

This configuration flexibility is achieved by exposing a parameter within our software architecture, accessible through the /sys file system. This level of configurability enables dynamic adjustments based on the system's operational needs and the criticality of the protected environment. By providing this flexibility, administrators can fine-tune the system's response, choosing between process termination or blocking based on the current circumstances and security requirements.

Our system implementes a Time-to-Live (TTL) mechanism as a form of rate-limiting control to prevent excessive page copies from being generated for the same user ID. Once an event (kill/block) occurs due to surpassing the threshold, the corresponding entry in the hash table associated with that user enters a TTL period.

During this TTL period, the counter within the hash table entry is decremented each time a copy of an executable page related to the same user ID is successfully delivered to the user-space agent. However, if processes from the same user generate additional page faults requesting page copies during this TTL period, they are still subjected to termination or blocking. Essentially, no new page copies are allowed for processes of that same user throughout the entire TTL duration.

Also, to contrast DoS caused by excessive memory usage of this second hash table, we added a second TTL. If the entry of the hash table associated with a given user ID keeps a counter that remains set to zero for the whole TTL, then this entry is removed from the hash table—it will be reinserted in the future upon a page fault interception for a program run from the same user ID. This avoids keeping useless entries in this hash table thus contrasting attacks that can be based on the exploitation of programs with the setuid flag active. In fact, these would allow an attacker to operate via multiple user ID values, enlarging at some point in time the number of entries of the hash table.

## 4 Assessment

For the assessment of our proposal, we concentrate on two key aspects. Firstly, we assess our system capability of detecting malware. Concerning this point, our emphasis lies on the ability of JITScanner to maintain effectiveness against both plain and packed malware. In relation to this aspect, we mention again that the Yara-based signature checker used in JITScanner is a kind of plugin that actually determines the final false positive/negative rate. However, at the same time, the core part of JITScanner is the kernel-level engine for enabling the creation of the snapshots of any page-image that is actually used by an application—in terms of instruction fetch. Hence, JITScanner can support the employment of differentiated signature checkers—including signature checkers that will be developed in the future—which can clearly provide improvements in terms of correct rate of identification of malware. Secondly, JITScanner is devised as a system that can offer continuous support for malware detection, being active on operating systems where regular (non-malware) applications are anyhow used. Hence, assessing the performance impact of JITScanner on common applications, as well as memory usage, is a relevant additional aspect we come with in this experimental study.

Table 2: Families of the samples used

| Family Name | Malware type | Number of samples |
|---|---|---|
| Emotet | Trojan | 2 |
| Mirai | Botnet | 52 |
| Tsunami | Botnet | 71 |
| Gafgyt | Trojan | 320 |
| XMRIG Miner | Coin-miner | 21 |

To evaluate the effectiveness and performance of our solution, a series of tests have been conducted on a VMWare virtual machine with Linux kernel version 5.13 as the guest operating system, utilizing a host with an i7-1165g7 2.80 GHz Intel processor with 8 cores, and 32GB of RAM.

## 4.1 Effectiveness

Our effectiveness testing focuses on two primary objectives:

1. Measuring the "signature flexibility" of JITScanner, which is the extent to which signatures from plain malware remain effective for a variant in the same family.

2. Measuring the "signature retention" of JITScanner, which is the extent to which signatures from plain malware remain effective for their packed counterparts.

For the malware dataset, we selected the "VirusShare_ELF_20200405," which is the latest dataset provided by the free malware sharing website VirusShare, comprising approximately 40,000 samples [25]. To showcase the effectiveness of JITScanner, and of a representative tool we used as competitor, namely ClamAV [1], we executed the samples both in their original state and after packing them with a simple packer sourced online [12]. From the original 40,000 we picked 1930 samples that where compatible with our chosen packer and scanned them with ClamAV. Out of these, we extracted 466 samples which ClamAV associated to an immediately identifiable malware family, these are reported in Table 2. After this, we run all samples, both in plain and in packed form, under JITScanner and logged the pages whose accesses in write-execute and/or execute mode have been intercepted[1].

For the logged pages with samples associated with a family, we chose the most common signature, ensuring that it did not yield any false positives during normal operations in our test. For the selection of such most common signature, we took from each logged page a sequence of bytes of code of length sufficient to identify actual machine instructions that typically appear after the prologue (e.g., classical push instructions at the beginning of a function) of the block of code where the thread jumped for fetching instructions into the page. For the different families of malware we analyzed, the length of this sequence of bytes was between 15 and 20. The most common signature of each family has been added to the Yara rules database. In particular, we added these signatures to the user agent of JITScanner, deploying it and reanalyzed all samples on our automated malware testing facility, namely PHOENIX [8]. For this effectiveness test, the meterpreter shellcode check on the executable page carried out at kernel level did not identify malware. Hence, malware identification is essentially carried out by the user level agent.

---
[1]All the datasets exploited in this study have been made available at: https://github.com/Capo80/Malware_Datasets

Table 3: Signature flexibility of JITScanner and ClamAV

|            | Family      | Detected Plain | Total Samples | Signature Flexibility |
|------------|-------------|----------------|---------------|-----------------------|
| JITScanner | Emotet      | 2              | 2             | **100%**              |
| ClamAV     | Emotet      | 1              | 2             | 50%                   |
| JITScanner | Tsunami     | 57             | 71            | **80.2%**             |
| ClamAV     | Tsunami     | 33             | 71            | 46.47%                |
| JITScanner | XMRIG_Miner | 13             | 21            | **61.90%**            |
| ClamAV     | XMRIG_Miner | 8              | 21            | 38.09%                |
| JITScanner | Gafgyt      | 76             | 320           | 23.75%                |
| ClamAV     | Gafgyt      | 238            | 320           | **74.75%**            |
| JITScanner | Mirai       | 15             | 52            | **28.84%**            |
| ClamAV     | Mirai       | 8              | 52            | 15.38%                |

To assess point 1, we decided to select the most common signature for each family, which we identified through our log of extracted pages. Subsequently, we compared the efficacy of these signatures against the most common signature detected by ClamAV. The results are presented in Table 3. As we can see, JITScanner has a comparable or superior effectiveness across all analyzed families, with the exception of Gafgyt. This discrepancy is likely attributable to the limited number of pages per sample captured in our logs for Gafgyt—the average is 4—which is significantly lower compared to other families, this prevented us from finding an effective signature. We believe that this type of malware detects that it is running on a testing environment (e.g., because of the detection of a reduced number of CPUs) and gives rise to a rapid termination. Such short or failed executions can lead not to intercept actual activities that the malware can (at least potentially) carry out via WX pages that include code signatures—which would have been observed by JITScanner. However, at the same time this is the scenario where the malware becomes essentially idle, not really leading to real security problems.

To assess point 2, we compared the traces extracted from the plain malware to those obtained from the packed samples. It is clear that all signatures are retained if the pages extracted from the plain sample constitute a subset of those extracted from the packed sample. The packed samples will obviously execute over more pages, as they need to decrypt the payload from memory. The details of the experiments are summarized in Table 4. As observed, there exists a substantial difference between the effectiveness of ClamAV and JITScanner when dealing with packed malware. ClamAV lacks a mechanism for detecting signatures at runtime. This makes it not capable to effectively detect packed samples once their signature is unpacked in memory for its actual usage. Conversely, JITScanner is able to use the same signatures for the most of all samples examined, recognizing the 75.91% of them as malware. Also, some malwares are no longer recognized because either the unpacker did not successfully lead to the execution of the malware code (hence JITScanner could not intercept the fetch of the malware instructions from memory) or the unpacking led to different access permissions for a few pages. As for the latter aspect, we found that some pages that are setup with read-execute permission with a regular loading of the ELF are instead setup as read-only by the unpacker. In this case, JITScanner does not perform any snapshot of the read-only page for a final check of its content when it is materialized—such snapshot is instead done for read-execute pages, and this gives rise to a different signature characterizing the application.

Table 4: Signature retention of JITScanner and ClamAV

|            | Sign. in Plain | Sign. in Packed | Total Samples | Signature retention |
|------------|----------------|-----------------|---------------|---------------------|
| ClamAV     | 515            | 0               | 515           | 0%                  |
| JITScanner | 515            | 391             | 515           | 75.91%              |

## 4.2 Performance

The second test we conducted was focused on performance evaluation. In particular, we carried out two distinct experiments in our study. Initially, we measured the performance of our solution when using conventional applications. Subsequently, we assessed the slowdown introduced by our solution for applications employing Just-in-Time (JIT) compilation, which make (important) usage of WX pages.

For the first test, we executed multiple trials of several classic Linux command-line utilities, both with and without the presence of our LKM, and with and without the synchronous kernel-level check. Each data point reports the average execution time over 2,000 samples. The results are shown in Figure 5. Additionally, the slowdown caused by our module without the synchronous check is reported in Table 5. Our findings indicate a slowdown in the range of 7% to 13%, which we can consider acceptable.

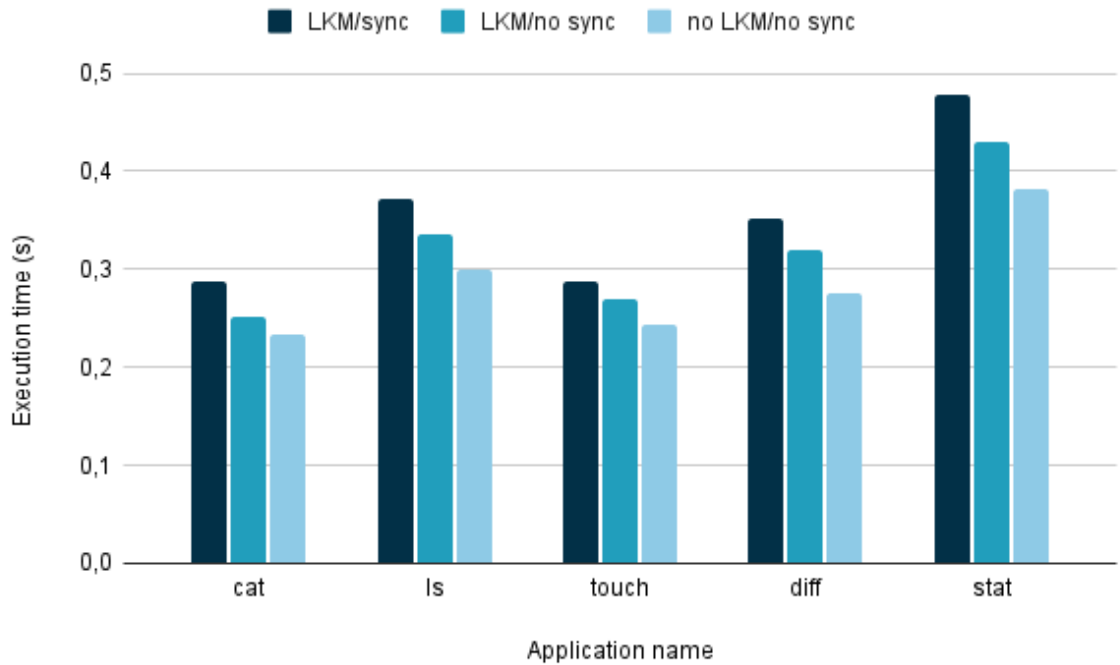

Figure 4: Execution time of common command-line utilities

For our second test, we selected three programming languages that are compatible with Just-in-Time (JIT) compilation: Lua, PHP, and Ruby; and used them to run a set of commonly utilized tests from the Debian benchmark game [16]. Also, to establish a performance baseline, we included the C language as a reference point. To mitigate excessive variability in our measurements, we configured the benchmark parameters to execute computationally non-trivial workloads, each with an execution time of approximately one second. Subsequently, similar to what done for the previous test, we conducted 500 iterations of the benchmarks under the following conditions: with and without the incorporation of our LKM, and with and without the synchronous level check of the page content.

The results of these experiments are presented in Figure 7 and the performance slowdowns are reported in Table 3. Our observations are as follows:

1. in the absence of the synchronous check, no test exhibited a slowdown exceeding 5%, and in some instances, no discernible slowdown was observed at all;

2. with the synchronous check enabled, the majority of tests experienced a slowdown of less than 10%, with some exceptions noted for the PHP language-based tests.

With these results and taking into account the possibility to disable the synchronous check, which, in certain scenarios, may have a greater impact on the slowdown, our architecture introduces a negligible performance overhead on JIT-based applications.

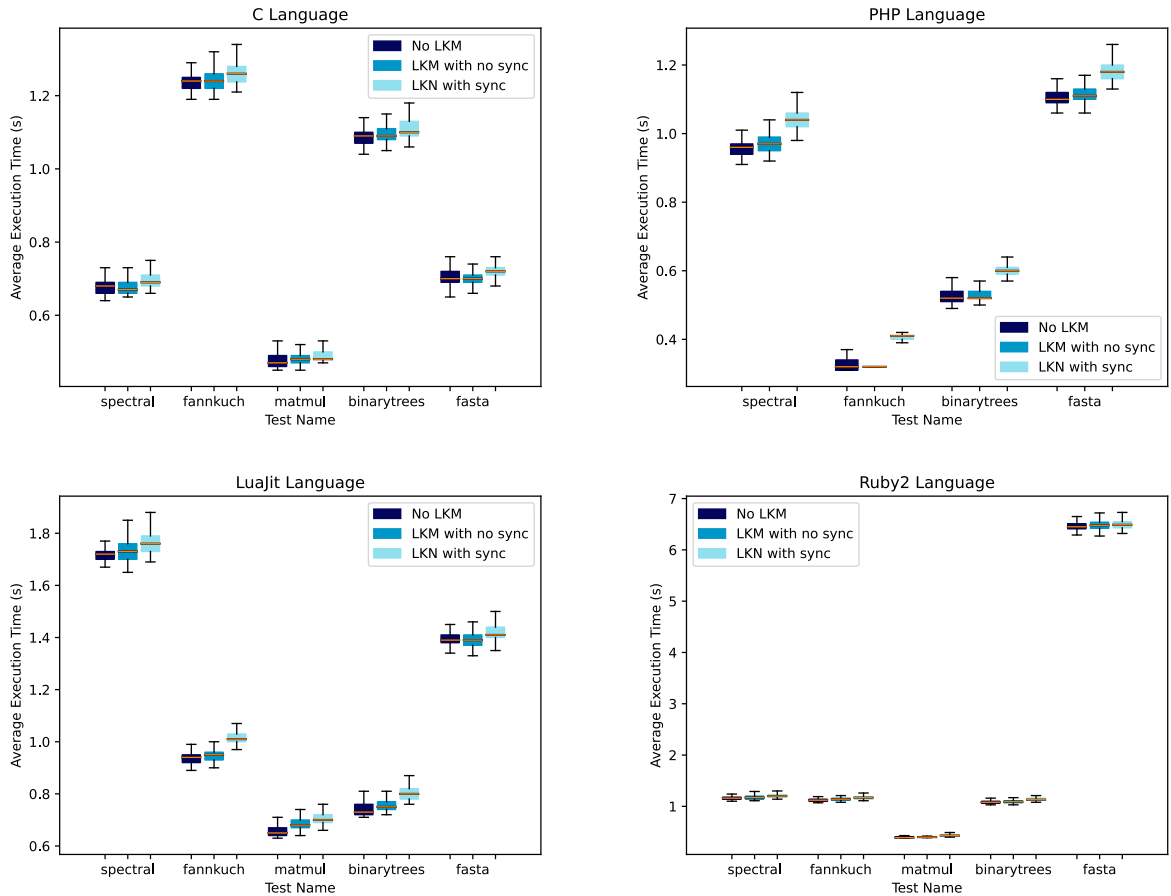

Figure 5: Execution time of JIT benchmarks

Additionally, it should be noted that our performance evaluation results refer to a kind of worst-case scenario for our system, as we tested it on very short-lived applications. In these cases, the cost of the initial security check on a page may not be fully offset by its subsequent accesses for instruction fetch. The slowdown is therefore expected to be less pronounced in applications where the longer overall execution time will mitigate the impact of the initial page check.

Table 5: Slowdown of common command-line applications with no synchronous page check

| Application Name | Slowdown |
|---|---|
| cat | 7.46% |
| ls | 10.57% |
| touch | 9.95% |
| diff | 13.90% |
| stat | 11.06% |

Table 6: Slowdown of JIT benchmark applications

| Language | Benchmark Name | No Sync Slowdown | Sync Slowdown |
|---|---|---|---|
| C | spectralnorm | 0.00% | 1.47% |
| C | fannkuchredux | 0.00% | 1.61% |
| C | matmul | 2.13% | 2.13% |
| C | binarytrees | 0.00% | 0.92% |
| C | fasta | 0.00% | 2.86% |
| LuaJit | spectralnorm | 0.58% | 2.33% |
| LuaJit | fannkuchredux | 1.06% | 7.45% |
| LuaJit | matmul | 4.62% | 7.69% |
| LuaJit | binarytrees | 2.74% | 9.59% |
| LuaJit | fasta | 0.00% | 1.44% |
| PHP | spectralnorm | 1.04% | 8.33% |
| PHP | fannkuchredux | 0.00% | 28.12% |
| PHP | binarytrees | 0.00% | 15.38% |
| PHP | fasta | 0.91% | 7.27% |
| Ruby2 | spectralnorm | 0.86% | 3.45% |
| Ruby2 | fannkuchredux | 1.79% | 4.46% |
| Ruby2 | matmul | 2.56% | 10.26% |
| Ruby2 | binarytrees | 0.93% | 4.63% |
| Ruby2 | fasta | 0.47% | 0.62% |

## 4.3 Memory Usage

The final test we performed was aimed at measuring the memory utilization of our system under normal conditions. We simulated a typical user interaction with software applications while monitoring the quantity of page snapshots awaiting analysis, we report our findings in Figure 6:

- From second 0 to second 60, our user was interacting with the web browser, checking his Facebook notifications, then moving to youtube to start watching a video. For the rest of the test the video kept playing in the background, while the user continued with his normal/foreground activity;

- From second 60 to 80, the user played a game of Mine Sweeper;

- From second 80 to 100, the user modified and saved a LibreOffice document;

- From second 100 to 130, the user opened Gimp and tested various options.

It can be observed that the vast majority of page snapshot creations (and their transfer to the user agent) occur at the beginning of the programs and when programs like web browsers start new processes, resulting in extremely small memory usage spikes; after the application has started the number of page transfers decreases sharply resulting in a near zero extra memory usage.

Furthermore, it is important to note that the test was conducted with a limited number of threads on the user side, due to the limitations of the used virtual machine. As a result, the software did not massively utilize our hash table supporting concurrent accesses, whose deep exploitation would further reduce memory usage and improve the overall performance.

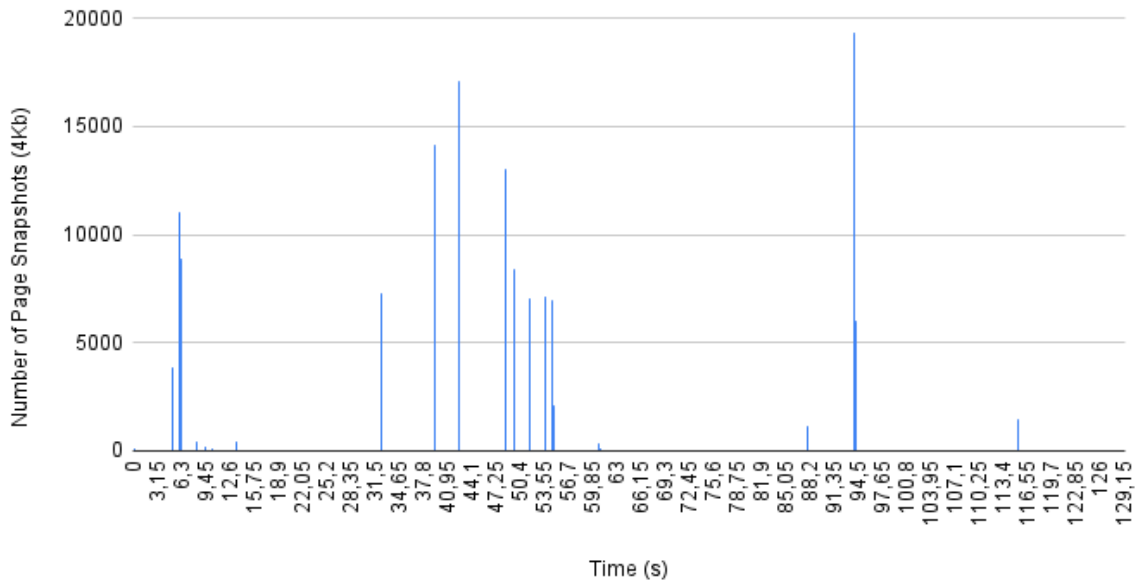

Figure 6: Memory usage of JITScanner

## 5 Usage models and potential extensions

JITScanner presents a flexible and adaptable solution to bolster the security of executable pages through its dynamic and just-in-time scanning methodology. The system's approach involves conducting security checks on binary code precisely at the instant when instructions are fetched from the most recent materialized page content. This strategy serves as an effective measure to mitigate potential risks posed by delayed execution of malicious code and malware that selectively decrypts malicious stages during runtime.

By focusing on scanning code at the moment of instruction fetching, JITScanner significantly reduces the window of opportunity for malicious activities to occur undetected. This proactive scanning mechanism enhances the system's ability to identify and neutralize threats that might attempt to evade detection by decrypting or executing malicious code after being loaded into memory. Overall, JITScanner's dynamic and real-time security checks contribute significantly to fortifying the protection of executable pages against various sophisticated threats and attack vectors.

**Flexibility and performance trade-offs.**
It is crucial to highlight that JITScanner does not enforce any specific binary code checking methodology. The current Proof-of-Concept (PoC) implementation showcases the use of Yara rules for asynchronous checks and the meterpreter shellcode signature search for synchronous checks merely as examples. However, in principle, JITScanner remains flexible to integrate *any* alternative technique capable of operating on a selected section of code, specifically the executable page accessed during instruction fetching.

In scenarios where a particular analysis method is resource-intensive or time-consuming, JITScanner has the capability to delegate its operation entirely to an external user-space thread. This delegation facilitates asynchronous analysis without impeding the execution of the code under examination. Additionally, for less critical services, JITScanner allows the bypassing of synchronous checks. This prioritizes application performance and accommodates potential delays in identifying malicious signatures, especially when the asynchronous check is activated. This adaptability ensures that JITScanner can accommodate various analysis methods while catering to diverse performance and security needs based on specific application contexts.

**Process management strategies.**

In our Proof-of-Concept (PoC) implementation, we have opted for simplicity by terminating the application upon identifying any security-critical aspect, thereby preventing potential further harm. However, JITScanner maintains the flexibility to employ various strategies, including adaptive approaches, based on specific circumstances. For instance, the system can initiate deeper monitoring only when specific content identified in a materialized executable page, accessed during instruction fetching, triggers a match against known malicious signatures.

As an example of adaptive strategies, JITScanner can dynamically activate the logging of system calls for a process that initiates execution using an executable page with identified malicious content. Additionally, for the same process, certain system calls could be selectively rejected at the kernel level, particularly considering input parameters (like file names) received by those system calls. This approach provides a granular control mechanism to prevent potentially harmful operations.

Moreover, as previously mentioned, JITScanner is designed to integrate seamlessly with external systems. When JITScanner detects malicious signatures actively used during execution, it can communicate with and trigger other external security systems to establish additional security barriers for the traced processes. This collaborative functionality enables JITScanner to synergize with external security measures, enhancing the overall security posture by creating a multi-layered defense against potential threats.

# 6 Conclusions

In this paper, we have introduced an innovative approach to dynamic executable analysis, focusing on the detection of malicious signatures within executable virtual pages. Our proposed approach is implemented in JITScanner, a Linux-oriented package built as a Loadable Kernel Module (LKM). JITScanner facilitates the inspection of any executable page snapshot each time its updated content in RAM is accessed for instruction fetching. This capability enables the management of pages with Write-Execute (WX) permissions, even when these pages undergo multiple updates throughout the application's lifecycle.

JITScanner is designed to support both synchronous and asynchronous page checks, offering configurability to meet the specific requirements of diverse environments. The combination of synchronous checks within the critical execution path and asynchronous checks performed outside this critical path ensures flexibility and adaptability in balancing security and system performance.

Our experimental data demonstrate the efficacy of JITScanner, showcasing its potential as an effective and scalable method for detecting and analyzing malware. This tool equips security professionals and researchers with a robust solution to identify malicious signatures within executable pages, contributing to the enhancement of system security and threat analysis capabilities.

The ability to dynamically perform comprehensive security checks on executable pages places JITScanner in a somewhat intermediate position between classical static analysis and behavioral analysis tools, and we believe it represents an additional tool to include in the arsenal of Blue teams.

# Acknowledgments

This project was partially supported by ECS Rome Technopole CUP N.: E83C22003240001.